\documentclass[sigconf]{acmart}
\AtBeginDocument{%
  }

\copyrightyear{2026}
\acmYear{2026}
\setcopyright{cc}
\setcctype{by-nc-nd}
\acmConference[UIST '26]{The 39th Annual ACM Symposium on User Interface Software and Technology}{November 02--05, 2026}{Detroit, MI, USA}
\acmBooktitle{The 39th Annual ACM Symposium on User Interface Software and Technology (UIST '26), November 02--05, 2026, Detroit, MI, USA}
\acmDOI{10.1145/3830398.3830563}
\acmISBN{979-8-4007-2856-3/2026/11}

\usepackage{cleveref}
\usepackage{enumitem}
\begin{document}

\title{VeriForge: Mitigating Latent Knowledge Gaps in Narrative Drafting via Mixed-Initiative Scaffolding}

\author{Ruqi Sun}
\email{sunrq2024@mail.sustech.edu.cn}
\affiliation{%
  \department{Department of Computer Science and Engineering}
  \institution{Southern University of Science and Technology}
  \city{Shenzhen}
  \state{Guangdong}
  \country{China}
}

\author{Jiaping Li}
\email{lijp2024@mail.sustech.edu.cn}
\affiliation{%
  \department{Department of Computer Science and Engineering}
  \institution{Southern University of Science and Technology}
  \city{Shenzhen}
  \state{Guangdong}
  \country{China}
}

\author{Wenhui Tao}
\email{wenhui9703@gmail.com}
\affiliation{%
  \department{Department of Computer Science and Engineering}
  \institution{Southern University of Science and Technology}
  \city{Shenzhen}
  \state{Guangdong}
  \country{China}
}

\author{Ximing Zheng}
\email{12311011@mail.sustech.edu.cn}
\affiliation{%
  \department{Department of Computer Science and Engineering}
  \institution{Southern University of Science and Technology}
  \city{Shenzhen}
  \state{Guangdong}
  \country{China}
}

\author{Yuefeng Tan}
\email{12410707@mail.sustech.edu.cn}
\affiliation{%
  \department{Department of Computer Science and Engineering}
  \institution{Southern University of Science and Technology}
  \city{Shenzhen}
  \state{Guangdong}
  \country{China}
}

\author{Jiahao Wei}
\email{12313319@mail.sustech.edu.cn}
\affiliation{%
  \department{Department of Computer Science and Engineering}
  \institution{Southern University of Science and Technology}
  \city{Shenzhen}
  \state{Guangdong}
  \country{China}
}

\author{Yuxin Ma}
\authornote{Corresponding author.}
\email{mayx@sustech.edu.cn}
\affiliation{%
  \department{Department of Computer Science and Engineering}
  \institution{Southern University of Science and Technology}
  \city{Shenzhen}
  \state{Guangdong}
  \country{China}
}

\renewcommand{\shortauthors}{Sun et al.}


\begin{abstract}
  Great fiction earns its verisimilitude through precise details, from how a longsword is gripped to pierce armor gaps to why a bleeding corpse cannot yet smell of decay, weaving domain expertise into the fabric of invented worlds. Current AI writing tools offer limited support for discovering and integrating unfamiliar domain knowledge into narrative. They require explicit queries that authors cannot formulate, generate finished prose that risks homogenizing voice, or assist only within the boundaries of what authors already know. We argue that AI should reveal latent knowledge gaps to writers while preserving their agency to transform discovered knowledge into authentic prose. Grounded in formative interviews with 9 fiction writers, we present VeriForge, a mixed-initiative writing system that divides cognitive labor so that the system assumes initiative over domain discovery while the author retains full initiative over narrative synthesis. VeriForge realizes this through three complementary mechanisms. Proactive inline highlighting flags potential knowledge gaps as authors draft. Dual-stream querying pairs conversational responses with source-anchored Knowledge Cards for direct fact extraction. A spatial Knowledge Canvas allows authors to organize and connect discovered knowledge across their writing. These mechanisms are powered by a graph-based retrieval-augmented generation pipeline grounded in domain-specific source materials. A within-subjects user study (N=12) provides preliminary evidence that this paradigm helps authors recognize previously overlooked knowledge gaps, supports creative exploration, and is perceived by expert raters to produce passages with stronger domain grounding in a controlled cold-start writing task.
\end{abstract}

\ccsdesc[500]{Human-centered computing~Interactive systems and tools}
\ccsdesc[300]{Human-centered computing~User studies}
\ccsdesc[300]{Computing methodologies~Natural language generation}
\keywords{AI-assisted writing, Mixed-initiative interaction, Narrative verisimilitude, Creativity support tools}
\begin{teaserfigure}
  \centering
  \vspace{-6pt}
  \includegraphics[width=0.96\textwidth]{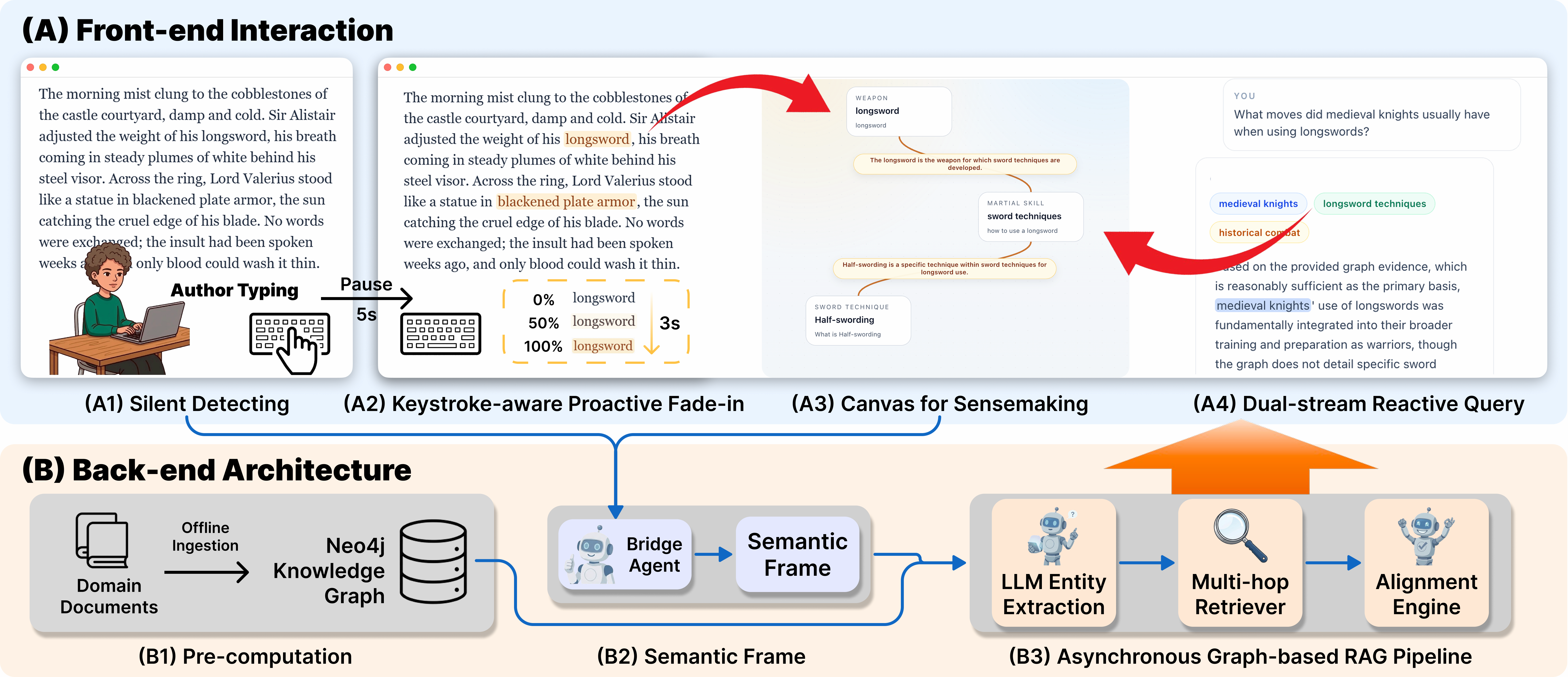}
  \vspace{-6pt}
  \caption{VeriForge overview. (A) Proactive highlights, Knowledge Cards, dual-stream queries, and a spatial canvas support drafting. (B) A Neo4j knowledge graph, Semantic Frame, and asynchronous Graph-based RAG pipeline ground these interactions.}
  \Description{A two-tier system diagram. The upper tier shows the author-facing workflow from left to right: an editor detects domain terms during typing, a five-second pause triggers a gradual inline highlight, selected terms become source-grounded cards and connected canvas nodes, and a query panel returns color-linked text and cards. The lower tier shows uploaded domain documents being ingested into a Neo4j knowledge graph, followed by a Bridge Agent and Semantic Frame that feed an asynchronous pipeline for entity extraction, multi-hop retrieval, and source alignment back to the interface.}
  \label{fig:teaser}
\end{teaserfigure}


\maketitle

\section{Introduction}

In creative writing, an immersive fictional world rests on the author's ability to establish \textit{verisimilitude}, a foundational illusion of truth~\cite{wolf2012building}, sustained through the careful weaving of domain-specific facts into prose~\cite{barthes1989reality, green2000role}. The \textit{knowledge-transforming} model of written composition~\cite{bereiter1987psychology} requires authors to navigate fluidly between a content space of factual knowledge and a rhetorical space of narrative craft. Established authors draw on internalized expertise to sustain this interplay; writers venturing into unfamiliar domains, however, face a more insidious obstacle. The \textit{Illusion of Explanatory Depth} (IOED)~\cite{rozensblit2002illusion} causes authors to overestimate their grasp of complex real-world mechanics while remaining entirely blind to the gaps in their understanding. The knowledge-transforming process breaks down not from a lack of effort, but from a lack of awareness.

Despite their unprecedented potential, large language models have so far failed to adequately support this knowledge-transforming process, and prior interaction paradigms each fall short in a distinct way. The dominant reactive paradigm, exemplified by conversational AI tools such as ChatGPT~\cite{chatgpt2026}, treats LLMs as on-demand reference tools, requiring authors to halt composition and formulate explicit queries~\cite{belkin1980anomalous}. Yet IOED ensures that the gaps most threatening to verisimilitude are precisely those authors cannot recognize, much less proactively investigate~\cite{rozensblit2002illusion}. Structural scaffolding tools help authors manage character networks and narrative timelines~\cite{park2026constella, masson2025visual}, but they primarily organize the author's existing knowledge rather than surfacing the external domain facts that verisimilitude demands~\cite{wolf2012building}. Direct text generation~\cite{yuan2022wordcraft, talaei2025storysage} attempts to fill this gap, but does not resolve the author's knowledge deficit; it simply bypasses it, delivering near-final prose that risks homogenizing authorial voice~\cite{anderson2024homogenization, agarwal2025ai, chakrabarty2024art} and displacing the cognitive engagement that knowledge-transforming requires~\cite{lehmann2023mixed, varanasi2025ai}. Across all three paradigms, the same fundamental need goes unmet: a mechanism that proactively presents the domain knowledge authors do not know they are missing, while preserving their agency to transform it into authentic prose~\cite{chen2025need, pu2025assistance}.

To ground our design empirically, we conducted formative interviews with 9 fiction writers (5 professionals, 4 hobbyists), revealing three critical breakdowns. First, authors harbored invisible knowledge gaps they could not recognize: one participant depicted a bleeding corpse emitting a rotting stench, unaware that the two states are physiologically incompatible. Second, fragmented workflows imposed a double cost: context-switching broke creative flow, and even when relevant facts were found, integrating them into the evolving narrative remained unsupported. Third, authors widely distrusted AI-generated facts, yet several noted that facts traced to a specific primary source felt trustworthy enough to adopt directly, suggesting that provenance is the more effective lever for restoring trust.

We present VeriForge, a mixed-initiative writing environment that scaffolds knowledge-transforming by dividing cognitive labor: the system assumes initiative over domain discovery while the author retains full initiative over narrative synthesis (\cref{fig:teaser}). On the system side, proactive inline highlights flag potential knowledge gaps as authors draft. On the author side, a dual-stream query panel pairs conversational responses with structured Knowledge Cards for direct fact extraction, and a spatial Knowledge Canvas allows authors to organize and connect discovered knowledge across their writing. Underlying these mechanisms, a graph-based retrieval-augmented generation pipeline anchors every retrieved fact to its exact primary source, directly addressing the trust deficit identified in our formative study.

To evaluate VeriForge, we conducted a within-subjects user study with 12 participants (8 creative writing hobbyists and 4 online novelists) comparing the full system against a strengthened baseline that replicates writers' existing workflows in a unified interface. We combined interaction logs, validated surveys, semi-structured interviews, and a controlled deception probing epistemic trust, alongside blind expert ratings and objective coding of the resulting narratives. Overall, this work makes the following three contributions:

\begin{enumerate}[label=(\arabic*), leftmargin=*]
    \item \textbf{A formative study with 9 fiction writers} that identifies three design requirements grounded in observed cognitive friction, revealing how IOED-driven blind spots prevent authors from self-diagnosing the knowledge gaps that most threaten verisimilitude.
    \item \textbf{VeriForge, a mixed-initiative writing environment} that supports the knowledge-transforming process for fiction writers venturing into unfamiliar domains, coupling proactive gap detection, dual-stream querying, and spatial knowledge curation with a traceable Graph-based RAG pipeline.
    \item \textbf{A within-subjects user study} (N=12, with blind ratings by two external published authors) providing preliminary evidence that proactive epistemic scaffolding helps authors recognize latent knowledge blind spots, supports creative exploration, and improves domain grounding in resulting passages during cold-start drafting.
\end{enumerate}

\section{Related Work}
 
\subsection{AI-Assisted Writing: From Generative Automation to Internal Scaffolding}
 
Current AI-assisted writing research follows two primary paradigms: \textit{narrative automation} and \textit{internal structural scaffolding}. The first paradigm focuses on AI-led content creation, from ghostwriting assistants like Wordcraft~\cite{yuan2022wordcraft} and StorySage~\cite{talaei2025storysage} to ideation tools such as Luminate~\cite{suh2024luminate} and TaleStream~\cite{chou2023talestream}. Others explore richer modalities for steering generation, such as TaleBrush's~\cite{chung2022talebrush} line-sketch control of a protagonist's fortune arc, or retrieve external reference material like ScriptViz's~\cite{rao2024scriptviz} movie visuals for screenwriting. While effective for brainstorming, these systems operate on high-level synthesis of plots and ideas, significantly shortening the inference distance~\cite{anderson2024homogenization} between prompt and output and risking homogenization of authorial voice~\cite{anderson2024homogenization, agarwal2025ai, chakrabarty2024art}.
 
In response, a second paradigm has emerged that shifts AI from a generative leader to a human-led assistant, focusing on the internal logic and multimodal planning of a story. Recent systems like Constella~\cite{park2026constella} and Visual Story-Writing~\cite{masson2025visual} help authors maintain complex character networks and timelines, while Vistoria~\cite{fu2026vistoria} introduces instrumental operations to enable synchronized image-text co-editing. Scrivener similarly helps writers keep manuscript sections, notes, research files, corkboards, and outlines in one long-form writing workspace~\cite{scrivener2026overview}. Together, these tools organize material the author already has and prioritize authorial agency over generative automation.
 
However, a critical gap remains in both paradigms. Narrative automation targets the rhetorical delivery of ideas, while structural assistants organize the author's existing knowledge. Neither proactively surfaces unfamiliar, source-grounded domain knowledge or adequately supports the \textit{knowledge-transforming} process~\cite{bereiter1987psychology} required to synthesize unfamiliar, specialized domain facts into prose, the very process on which verisimilitude depends~\cite{wolf2012building}.

\subsection{Mixed-Initiative and Proactive Assistance}
Traditional reactive assistants fail in domain-specific writing because users often cannot articulate their information needs~\cite{belkin1980anomalous, wu2023inscit}. IOED makes this problem acute in creative writing: authors may be unaware of the boundaries of their own domain knowledge, meaning the gaps most threatening to verisimilitude are precisely those they cannot formulate as queries. While proactive agents in conversational search~\cite{mei2025interquest} and task support~\cite{arakawa2024prism} mitigate this, studies in AI-assisted programming caution that proactive generation causes workflow disruptions and diminishes users' sense of ownership~\cite{chen2025need, pu2025assistance, lehmann2023mixed}. Excessive system autonomy further erodes epistemic trust compared to suggestive strategies~\cite{kraus2020effects}. Wan et al.~\cite{wan2024secondmind} echo this in a study of prewriting with LLMs, finding that rigid allocation of control to either side hinders creative collaboration.

VeriForge draws on mixed-initiative principles~\cite{horvitz1999principles, fleming2001user} but applies them to a specific cognitive asymmetry: the system assumes initiative only over domain discovery, where IOED renders authors unable to act for themselves, while the author retains full initiative over narrative synthesis, where the cognitive work of transforming knowledge into one's own prose is itself the mechanism through which deeper understanding forms~\cite{arnold2017understanding, bereiter1987psychology, kang2023synergi, zheng2024disciplink}.
 
\subsection{Spatial Sensemaking and External Cognition}
Domain-specific fiction writing is a non-linear, information-intensive task that demands ontological consistency~\cite{wolf2012building}, placing a high knowledge-transforming load on authors that necessitates external cognitive support~\cite{bereiter1987psychology}. However, the physical separation between linear editors and spatial tools induces context-switching and resumption lag that fragments the very flow such tools aim to support~\cite{iqbal2008understanding}. While recent systems integrate canvases for character and plot management~\cite{park2026constella, masson2025visual}, and ClueCart~\cite{wang2025cluecart} demonstrates the value of structured spatial organization for reconstructing narratives from fragmented clues, they treat spatial layouts primarily as an output medium for human viewing~\cite{suh2024luminate, qin2025toward, suh2023sensecape}. Patchview~\cite{chung2024patchview} and composable prompting workspaces~\cite{amin2025composable} show that spatial interactions can steer LLM behavior and enhance perceived control over generation, but spatial reasoning still does not feed back into the retrieval process itself. Without awareness of what has already been curated, such systems risk producing retrievals that are redundant or contradictory to the world-logic under construction.
 
\subsection{Knowledge-Transforming and Domain-Specific RAG}

Growing research interest in domain-specific AI has shifted focus from general-purpose generation toward the synthesis of specialized expertise~\cite{bereiter1987psychology}, adapting LLMs to vertical fields such as agriculture~\cite{vizniuk2025comprehensive}, technical service~\cite{wulf2024exploring}, and ESG reporting~\cite{gupta2024knowledge} through domain-specific reading comprehension~\cite{cheng2024adapting} and culturally grounded platforms~\cite{team2025fanar}. However, traditional RAG~\cite{lewis2021retrieval} retrieves facts in isolation, failing to capture relational connections between concepts. GraphRAG addresses this by organizing knowledge as a graph structure, enabling retrieval that reflects how concepts relate to one another~\cite{edge2025local}, a capacity that has proven valuable in expert environments where isolated snippets fail to convey the full picture~\cite{irbe2025investigating}.

These approaches, however, are rarely designed with creative writers in mind. Retrieving a fact is only the first step; the harder challenge is weaving interconnected domain knowledge into authentic prose, and when retrieved facts carry no provenance, authors must verify them independently, breaking creative flow. Taken together, the literature reveals a consistent blind spot: no existing system simultaneously detects the knowledge gaps authors cannot self-diagnose, embeds discovery within the writing surface, supports spatial curation of what is found, and anchors every fact to its primary source. To ground these needs empirically before proposing a design, we conducted a formative study with fiction writers.

\section{Formative Study}

To understand information-seeking breakdowns during domain-specific fiction writing, we conducted artifact-driven interviews with 9 fiction writers (5 professionals, 4 hobbyists), recruited through social media platforms and compensated with the equivalent of 15 USD in local currency each. Participants' writing spanned diverse genres including folklore, science fiction, crime fiction, historical romance, and cosmic horror, ensuring breadth of domain-specific challenges. Demographics are provided in Appendix~\ref{app:demographics}.

Each 60-minute session was structured in two phases. In the first phase, we asked participants about their general writing practices, including how they research unfamiliar domains, what tools they use to organize reference material, and where they experience the greatest friction in this process. In the second phase, we conducted a retrospective walkthrough of a chapter or passage each participant identified as particularly research-intensive. Participants brought the relevant draft along with any notes or reference materials they had used. Before each session, the research team independently fact-checked the submitted passages using domain references and LLM-assisted verification. During the walkthrough, we guided participants through these passages, asking them to reconstruct their research process and reflect on specific details where our fact-checking had revealed potential inconsistencies. Two researchers independently coded the interview transcripts using thematic analysis, resolving disagreements through discussion until consensus was reached. The full interview protocol and codebook are provided in the supplemental material.

\subsection{Formative Study Findings}

\subsubsection{The Importance of Verisimilitude and Latent Knowledge Gaps}

All nine participants agreed that domain-accurate details are essential to sustaining reader immersion. F03 observed that while specialized knowledge may not be the centerpiece of a story, it serves as a crucial supporting layer that makes characters more believable and the reading experience more tangible. F07 recalled accidentally describing northern China's centralized winter heating as a switch-on air conditioner, noting that this single misplaced detail instantly broke reader immersion. Yet despite this shared commitment to verisimilitude, authors frequently harbored blind spots they could not recognize. F09 depicted a corpse that was both still bleeding and emitting a rotting stench, unaware that a body fresh enough to bleed would not yet have begun to decompose. F02 set a global vaccine rollout within three months, far beyond the medical infrastructure of her novel's historical period. These examples illustrate that the most dangerous gaps are not those authors choose to ignore, but those they never perceive in the first place. These gaps were not limited to historical or martial settings: they involved forensic timing, regional infrastructure, and medical logistics.

\subsubsection{Fragmented Workflows and the Cost of Creative Disruption}

When authors do attempt to fill knowledge gaps through external research, the cost is qualitatively different from task-switching in other knowledge work. Our participants described interruptions not as recoverable resumption lags but as risking the irreversible loss of an emotional and imaginative state. F03 characterized emotional continuity as a non-renewable resource: \textit{``Even though I know the direction, I won't be able to recapture the feeling I had today.''} F09 noted that the research process itself could contaminate creative intent, as encountering tangential information would break her \textit{``original vision and outline.''} Even when relevant information was found, translating it into narrative prose remained unsupported: F01 observed that factual references \textit{``simply cannot support the functions I need for my plot''} without substantial reinterpretation.

\subsubsection{Trust Deficit and the Fragility of Verification}

AI tools offer a potential shortcut past this research burden. However, our interviews revealed two concerns more specific to creative writing. First, authors resisted AI-generated prose not merely for its inaccuracy but because it threatened their voice. F04 described generic AI vocabulary as carrying a \textit{``plastic texture,''} and F01 warned that seeing detailed AI-generated text caused her writing to \textit{``inevitably drift toward someone else's words.''} Second, several participants noted that when facts were explicitly traced to a specific primary source, they felt confident enough to use them directly without independent verification. This suggested that provenance rather than output quality is the more effective lever for restoring trust, and motivated our decision to present domain knowledge as structured raw material rather than finished prose.

\subsection{Design Goals}

Drawing on these three breakdowns, we formulated four design goals (DGs) to guide VeriForge. Findings 1 and 3 each map onto a single DG, while Finding 2 yields two complementary goals addressing its distinct dimensions of context switching and knowledge curation.

\begin{itemize}[leftmargin=*]

\item \textbf{DG1: Proactively Illuminate Latent Knowledge Gaps.} Since authors cannot prompt for constraints they do not realize exist \textbf{(Finding 1)}, the system must act as a silent background collaborator, continuously monitoring drafts to identify potential knowledge gaps before they are cemented into the narrative.
\item \textbf{DG2: Preserve Creative Flow via In-Situ Integration.} To reduce resumption lag and protect authorial focus \textbf{(Finding 2)}, contextual cues and search mechanisms must be embedded directly into the primary editor, shielding authors from the attention hijacking of external tools.
\item \textbf{DG3: Support Fluid, Low-Overhead Knowledge Curation.} To accommodate ad-hoc writing styles and reduce curation friction \textbf{(Finding 2)}, the system must make it easy for authors to capture, organize, and reconnect discovered domain details without imposing a rigid organizational scheme.
\item \textbf{DG4: Restore Trust via Structured, Source-Traceable Knowledge.} Since authors distrust AI-generated facts and resist finished prose as a threat to their voice \textbf{(Finding 3)}, retrieved knowledge must be presented as structured raw material linked to its exact source passage, preserving both verifiability and authorial ownership.
\end{itemize}

\section{System Design}

Guided by our formative study, VeriForge is a mixed-initiative writing environment featuring a dual-layer architecture: a front-end coupling a proactive editor with a spatial canvas (\textbf{DG1-3}), powered by a graph-based retrieval back-end anchored in curated literature (\textbf{DG4}) (\cref{fig:teaser}).

\begin{figure*}[h]
  \centering
  \includegraphics[width=0.9\linewidth]{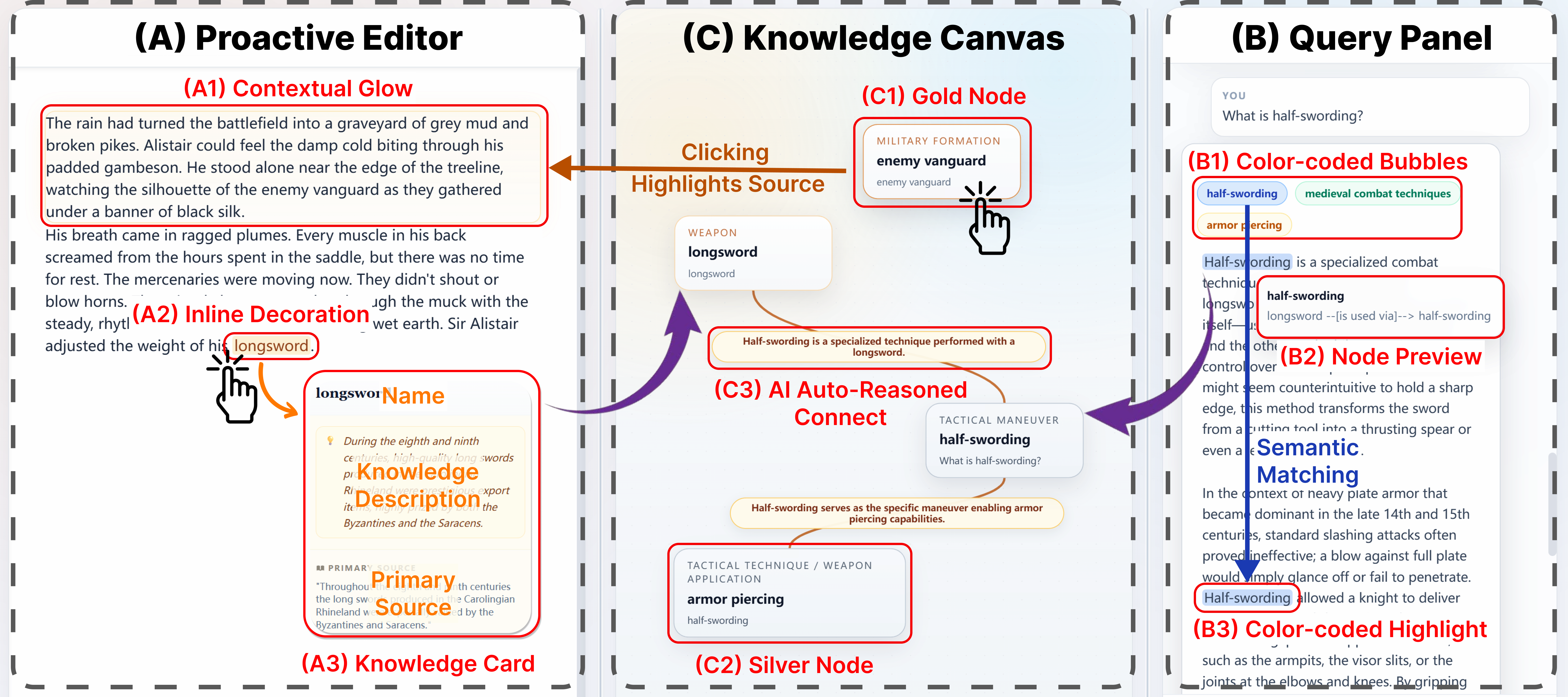}
  \caption{The VeriForge mixed-initiative front-end. (A)~The Proactive Editor flags domain concepts as inline highlights (A2) that expand into source-anchored Knowledge Cards (A3); clicking a gold Canvas node glows its source passage (A1). (B)~The Query Panel pairs color-coded Knowledge Card bubbles (B1) with a conversational text stream via semantic color-matching (B3); hovering reveals graph provenance (B2). (C)~The Knowledge Canvas organizes gold nodes from highlights (C1) and silver nodes from queries (C2), connected by AI Auto-Reasoned edges (C3).}
  \Description{Three side-by-side interface views linked by arrows. The left editor highlights ``longsword'' in a draft and opens a card containing an AI description and a primary-source excerpt. The center canvas contains gold nodes created from editor highlights and silver nodes created from queries, with labeled AI-generated edges; selecting a gold node points back to its source passage. The right query panel pairs colored concepts in a conversational response with matching card bubbles and previews their graph provenance.}
  \label{fig:system}
\end{figure*}

\subsection{System Overview and User Walkthrough}

To illustrate VeriForge's workflow, consider Marie, a novelist drafting a medieval duel. She begins by uploading several reference books on European martial arts; the system automatically ingests them into a domain knowledge graph. As she types (\cref{fig:system}, A), the system silently monitors her prose and detects \textit{longsword} as a domain concept with relational complexity in the knowledge graph. After a 5-second keystroke pause, the word fades into a subtle inline highlight (A2). Marie clicks it, and a Knowledge Card expands in situ (A3). The card separates an AI-generated description from a quoted source passage. The description explains why the concept matters in the current writing context. The source passage provides the supporting evidence.

Needing more concrete techniques, Marie turns to the Query Panel (\cref{fig:system}, B) and asks \textit{``What is half-swording?''} The dual-stream response simultaneously produces a conversational text stream and color-coded Knowledge Card bubbles (B1). Matching colors couple highlighted concepts in the text to their corresponding cards (B3), and hovering over a bubble reveals its graph provenance (B2).

Marie adds the retrieved nodes to her Knowledge Canvas (\cref{fig:system}, C). Gold nodes sourced from editor highlights (C1) and silver nodes from her query (C2) populate the workspace. \textit{AI Auto-Reasoned Connect} immediately draws labeled edges between them (C3), surfacing that half-swording is a specialized technique performed with a longsword. Clicking the gold \textit{enemy vanguard} node glows its originating passage back in the editor (A1), maintaining full provenance traceability throughout her writing session.

\subsection{Mixed-Initiative Front-End (DG1-3)}

\subsubsection{Flow-Aware Proactive Highlighting (DG1 \& DG2)}

\begin{figure}[h]
  \centering
  \includegraphics[width=\linewidth]{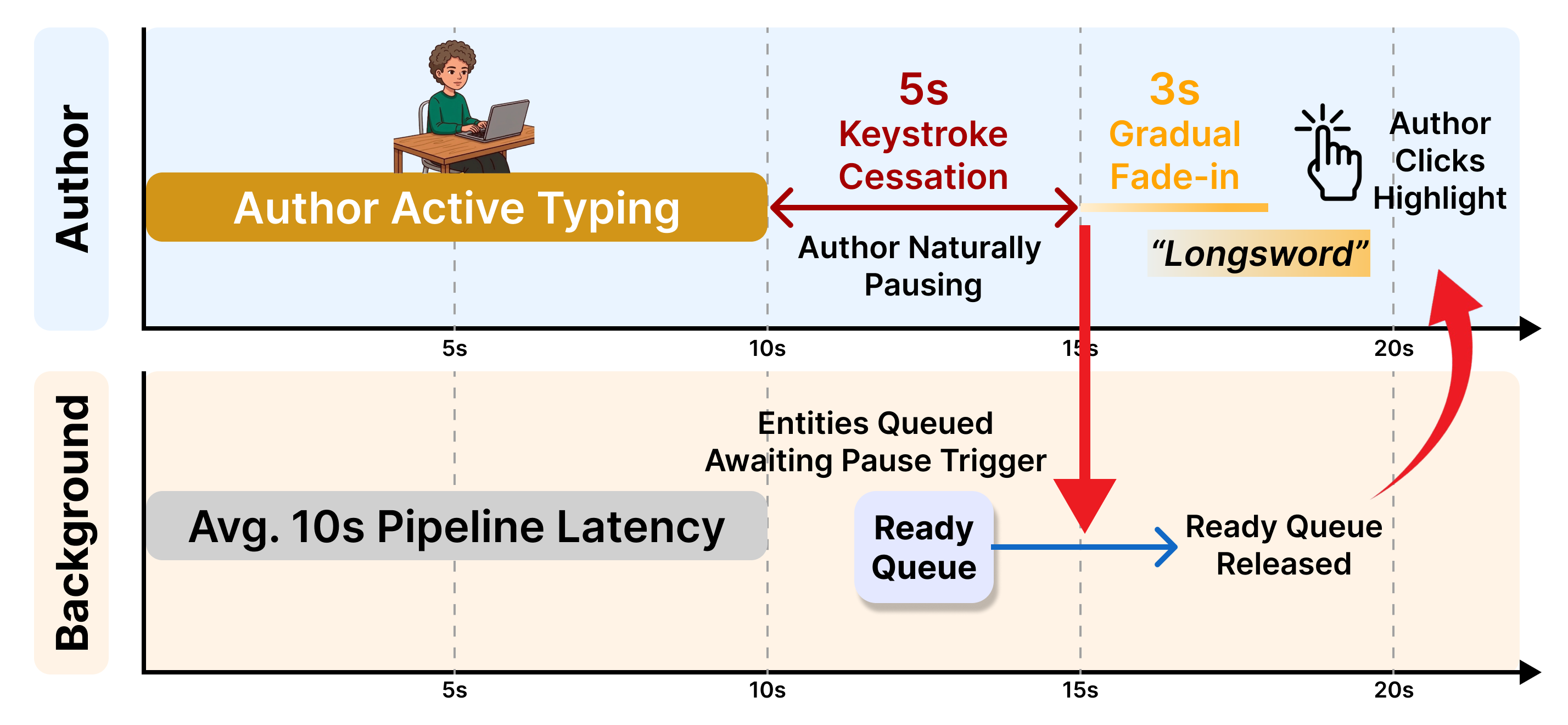}
  \caption{Flow-aware proactive highlighting. Asynchronously retrieved entities wait in a ready queue and fade in only after a 5-second keystroke pause, preserving author flow.}
  \Description{Two synchronized timelines from zero to twenty seconds. The upper timeline shows continuous author typing, a natural five-second pause, a three-second fade-in of the ``longsword'' highlight, and a click. The lower timeline shows an approximately ten-second retrieval pipeline running during typing, entities waiting in a ready queue, and the queue being released only after the pause trigger.}
  \label{fig:highlight}
\end{figure}

A key design assumption follows directly from IOED: authors venturing into an unfamiliar domain often use familiar terms (e.g., longsword, armor) without recognizing the constraints those terms carry. VeriForge therefore does not determine whether a term is used correctly before highlighting it. It surfaces candidate knowledge gaps: source-backed terms in the author's own prose whose graph neighborhoods suggest useful domain knowledge to inspect. Uploading source material reflects subject-level awareness, not gap-level awareness; authors may know a medieval combat scene needs references without knowing which terms or relations will matter while drafting.

Selection proceeds in three steps. First, the current prose span is embedded, matched against the entity vector index, and expanded by one hop in the graph. Second, an Alignment Engine proposes annotations grounded in that subgraph. Candidates are kept only when their trigger phrase appears in the author's text, links to a source-backed entity or relation, and is not merely a generic scene word. Here, relational complexity means source-backed relations likely to affect narrative action or interpretation, such as causal consequences, required tools, risks, preventions, implications, or cultural constraints. Third, the surviving candidates enter a ready queue. Edits overwrite pending requests; highlights fade in only after a 5-second keystroke pause (\cref{fig:highlight}). Clicking a highlight opens a source-anchored Knowledge Card (A3) in the editor. Thus, a highlight is a low-cost hypothesis about knowledge value, not a claim that the author has made an error.

\begin{figure*}[h]
  \centering
  \includegraphics[width=\linewidth]{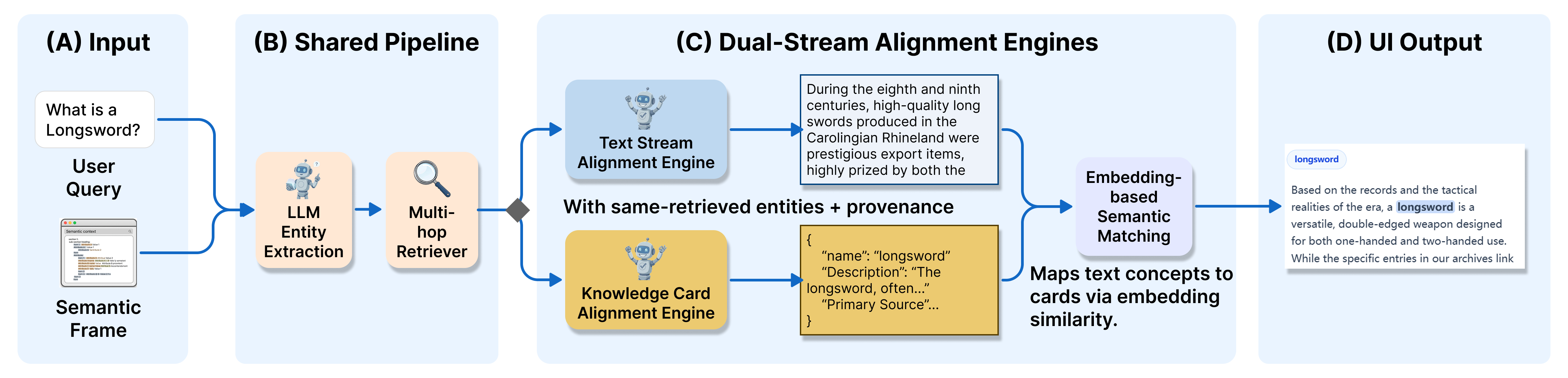} 
  \caption{The dual-stream query pipeline. A query and Semantic Frame are processed through entity extraction and multi-hop retrieval (B), then routed to two parallel Alignment Engines (C) producing a text stream and structured Knowledge Cards. Semantic matching (D) couples highlighted concepts in the text stream to their corresponding Cards.}
  \Description{A left-to-right pipeline with four stages. A user query and the current Semantic Frame enter shared LLM entity extraction and multi-hop retrieval. The same retrieved entities and provenance then split into a text-stream Alignment Engine and a Knowledge Card Alignment Engine. Their outputs are a conversational passage and a structured card. An embedding-based matcher links concepts in the passage to the corresponding card for color-coded display.}
  \label{fig:dual}
\end{figure*}

\subsubsection{Dual-Stream Reactive Querying}

To balance conversational familiarity with structured sensemaking, VeriForge's query panel employs a dual-stream approach (\cref{fig:dual}). A single retrieval result is routed to two parallel Alignment Engines: one produces a conversational text stream, while the other simultaneously packages the same retrieved entities into structured Knowledge Cards rendered as color-coded bubbles (\cref{fig:system}, B1). Embedding-based semantic matching couples highlighted concepts in the text stream to their corresponding bubbles via shared colors (B3), creating a tight visual link between explanation and evidence. Hovering over a bubble reveals its graph provenance (B2). This design encourages authors to extract verified facts directly to their canvas as raw ingredients for active synthesis, rather than passively consuming finished prose.

\subsubsection{Fluid Spatial Sensemaking (DG3)}

The Knowledge Canvas functions as a unified spatial workspace where insights from both the editor and the query panel converge (\cref{fig:system}, C). Knowledge Cards added from editor highlights become gold nodes (C1), while those added from queries become silver nodes (C2), maintaining a visual record of each fact's provenance. Both retain the card's AI-generated description and source passage. Gold nodes also link to the prose span that triggered the highlight. The link is stored against this span rather than the bare trigger term, so it remains available if the term is later edited. Silver nodes link to the relevant query turn. \textit{AI Auto-Reasoned Connect} automatically calculates and visualizes semantic connections between nodes as labeled edges (C3). Clicking a gold node triggers a contextual glow on its originating passage in the editor (\cref{fig:system}, A1), while clicking a silver node scrolls the query panel back to the relevant conversation turn. Critically, canvas interactions feed directly into the Semantic Frame that informs all subsequent retrievals, ensuring the system remains aware of what has already been curated.

\subsection{Traceable Graph-Based RAG Back-End (DG4)}

VeriForge adopts a graph-based architecture over standard flat-vector retrieval for three reasons. First, graph structures explicitly encode relationships between domain concepts rather than storing facts in isolation, enabling the system to surface not just what a concept is but how it connects to others~\cite{Pan_2024,edge2025local}. Second, binding provenance metadata to nodes and edges at ingestion time ensures every retrieved fact remains traceable to its exact source passage, a guarantee that chunk-based retrieval cannot provide by design~\cite{luo2025hyper}. Third, multi-hop graph traversal surfaces topologically adjacent concepts the author would not have thought to query, directly supporting proactive gap detection~\cite{edge2025local, irbe2025investigating,he2024gretrieverretrieval}. Drawing on these principles~\cite{edge2025local}, VeriForge implements a domain-specific knowledge graph pipeline tailored to creative writing's provenance and interactivity requirements; full architecture details and a technical evaluation against a standard RAG baseline are provided in the supplemental material.

The ingestion pipeline is fully automated: authors upload their domain texts and the system handles entity extraction, relationship inference, and provenance binding without manual configuration (\cref{fig:teaser}, B1), typically requiring 5 to 10 minutes per book. Every node retains explicit provenance, recording both the source document and the exact passage from which it was derived.

At runtime, an asynchronous Bridge Agent silently maintains a Semantic Frame capturing both the evolving prose and the author's canvas topology (\cref{fig:teaser}, B2), ensuring all retrievals reflect the full scope of what has been written and curated without blocking UI interactions. Triggered by a typing pause or an explicit query, the Semantic Frame passes through three phases (\cref{fig:teaser}, B3): LLM entity extraction identifies domain concepts for proactive highlighting; a hybrid multi-hop retriever locates an anchor node then expands to semantically related neighbors, powering AI Auto-Reasoned Connect; and an Alignment Engine couples each entity to its exact primary source metadata before any information reaches the UI.

\subsection{Implementation Details}

VeriForge is implemented as a decoupled web application. The front-end is built with Vue 3 and TypeScript~\cite{vuejs,typescript,vueflow}, leveraging TipTap/ProseMirror for keystroke-aware inline highlighting and Vue Flow for the canvas~\cite{tiptap,prosemirror}. The Python/FastAPI back-end manages asynchronous tasks via an in-process ThreadPoolExecutor with 4 concurrent workers~\cite{fastapi}. Domain knowledge is housed in Neo4j~\cite{neo4j2026}, supporting both native vector indexing and structural graph expansion. We use the Qwen3.5-Flash API for text generation and entity extraction~\cite{qwen35flash}, selected to balance the response speed required by proactive highlighting, output quality, and API cost for sustained real-time interaction, and BAAI/bge-m3 for text embeddings~\cite{bgem3_baai}. Additional implementation details and system prompts are provided in the supplemental material.

\section{User Study}
 
To evaluate how VeriForge influences the creative writing process, context-switching overhead, and authorial agency during domain-specific fiction writing, we conducted a remote, moderated within-subjects user study. The study was approved by our institution's ethics review board. Our evaluation was guided by four primary research questions (RQs):
 
\begin{itemize}[leftmargin=*]
\item \textbf{RQ1 (Latent Gaps):} How does proactive, flow-aware highlighting affect authors' ability to identify and address latent knowledge blind spots compared to a reactive baseline?
\item \textbf{RQ2 (Creativity \& Retention):} How does the integration of spatial sensemaking influence authors' creative exploration, perceived creativity support, and long-term knowledge retention?
\item \textbf{RQ3 (Trust \& Agency):} How do authors calibrate their trust in retrieved facts, and does VeriForge's structured presentation affect their perceived creative agency?
\item \textbf{RQ4 (Outcome Quality):} Does the mixed-initiative workflow help authors produce short passages with stronger domain grounding during cold-start drafting?
\end{itemize}

\subsection{Study Design \& Baseline Rationale}
 
We employed a within-subjects design with two writing domains: Medieval European Martial Arts (e.g., longsword techniques, knight hierarchies) and Republican-era Chinese Martial Arts (e.g., polearm combat, traditional dueling protocols). We counterbalanced the order of both writing domains and system conditions across participants to minimize learning effects.

A key design decision was the choice of baseline. Many participants in our formative study researched unfamiliar domains by switching between a word processor and a conversational AI tool such as ChatGPT, occasionally supplementing with a note-taking surface. This pattern is consistent with the user-study sample, where 10 of 12 participants reported prior experience with AI writing tools. We initially piloted this representative tool chain (MS Word + ChatGPT + Miro) with three participants. However, the quantitative gap between conditions was dominated by differences in retrieval quality rather than interaction design: participants encountered frequent hallucinations and insufficiently specific responses from ChatGPT that eroded trust entirely, making it impossible to isolate the effect of our front-end design choices. We therefore treated these sessions as formative and redesigned the baseline to avoid confounding retrieval quality with interaction design.

The resulting strengthened baseline replicates a unified writing workflow consisting of a text editor, a Miro-like spatial canvas, and a conversational query panel in a single three-panel interface, all connected to the same graph-based retrieval and Semantic Frame backend as VeriForge. It removes the interaction mechanisms central to VeriForge: proactive inline highlighting, highlight-derived Knowledge Cards, dual-stream text-card coupling, gold nodes from editor highlights, and AI-reasoned canvas connections. The baseline provided no direct action for saving query results as cards or nodes, although participants could copy query text into manually created canvas nodes. Table~\ref{tab:baseline-comparison} summarizes the shared infrastructure and interface differences. This ensures that any observed differences between conditions reflect interaction design choices rather than underlying retrieval quality. Crucially, this design is deliberately conservative: because the baseline enjoys the same high-quality, source-grounded retrieval as VeriForge, any observed advantage must be attributed to the interaction paradigm itself rather than to superior retrieval infrastructure. In practice, this means our results likely underestimate VeriForge's marginal benefit over the fragmented tool chains that fiction writers currently use.

\subsection{Participants}
 
We recruited 12 participants (8 creative writing hobbyists and 4 online novelists) through local and online writing communities. This was an AI-experienced sample: 10 of 12 reported prior experience with AI writing tools. To ensure the study measured unfamiliar domain knowledge discovery (\textbf{RQ1}), we screened participants for little prior knowledge of Medieval Historical European Martial Arts or Traditional Chinese Martial Arts. Full demographics are provided in Appendix~\ref{app:demographics}. Each participant was compensated with the equivalent of 15 USD in local currency.
 
\subsection{Task: Constrained Story Continuation}
 
We designed a story continuation task requiring participants to draft a narrative scene highly dependent on accurate historical details. Participants received a brief story prompt setting the scene and characters two hours before the experiment. They were instructed to outline the plot mentally but not to conduct any external background research prior to the session.
 
During the 30-minute writing phase, participants produced a focused snippet of approximately 250 to 400 words. This time constraint was designed to isolate the cold-start phase of domain knowledge discovery, where IOED-driven blind spots are most acute.
 
\subsection{Procedure}
 
The study consisted of a 90-minute moderated session followed by a delayed remote test one week later. The main session began with a 5-minute tutorial, after which participants completed two 30-minute writing sessions. To further mitigate participant response bias~\cite{Dell2012yours}, the experimenter framed the two conditions neutrally during onboarding as a \textit{complex mixed-initiative interface} and a \textit{simple free-form interface}, without indicating which was the target of evaluation. Participants were encouraged to evaluate each interface based on its actual utility for their writing.
 
After the writing tasks, participants completed post-task questionnaires followed by a 15-minute semi-structured interview. To probe verification behavior (\textbf{RQ3}), we designed a controlled deception. Under normal operating conditions, all Knowledge Cards are derived exclusively from user-uploaded primary sources via the graph-based retrieval pipeline, making such errors extremely unlikely to occur naturally. We replaced the source passage on the second card with an unrelated passage while keeping its AI-generated description unchanged. This created a visible source-card mismatch. Participants could detect it by comparing the two fields, without prior domain expertise. Replacing the description instead would test hallucination detection.
 
One week after the main session, participants completed an unanticipated delayed recall task, writing a short paragraph per domain (5 minutes each) about the concepts they remembered.
 
\subsection{Measures}
 
\subsubsection{Interaction Logs \& Surveys}
 
We logged all system interactions throughout each writing session, including the number of queries issued, canvas node counts, and the number of lines written. Post-task surveys addressed each RQ through validated instruments and targeted custom items. For \textbf{RQ2}, we used the Creativity Support Index (CSI), which covers Exploration, Expressiveness, Immersion, Enjoyment, Results Worth Effort, and Collaboration, and an adapted 17-item 0--100 slider sensemaking scale~\cite{Alsufiani2017towards} covering information structuring, insight and internalization, relational visibility, ambiguity reduction, and knowledge-gap bridging. For \textbf{RQ3}, we administered an 11-item 7-point custom scale covering blind-spot alerting, canvas coordination, perceived internalization, trust, and authorial autonomy, supplemented by the controlled deception experiment described in Section~5.4. The NASA Task Load Index (NASA-TLX) measured workload using the standard six dimensions and weighted total score. The System Usability Scale (SUS) provided a standard 10-item usability check. Given our sample size, we treat these quantitative metrics primarily as structural triangulation for the qualitative findings derived from audio-recorded interviews, which were independently transcribed and coded by two researchers using thematic analysis, with disagreements resolved through discussion.
 
\subsubsection{Delayed Recall}
 
To operationalize long-term knowledge retention (\textbf{RQ2}), two researchers independently coded distinct domain-specific terms in each original story snippet and one-week recall paragraph, including weapon and technique names, institutional roles, cultural concepts, material objects, and procedural constraints. We computed both raw recall count and retention rate, defined as the proportion of original-snippet domain terms that reappeared in recall (recalled terms / original terms). This ratio partially controls for differences in exploration volume: because VeriForge participants incorporated more domain terms into their stories, their denominators were larger, making the retention rate comparison more conservative rather than more favorable to VeriForge. That VeriForge still yielded a higher retention rate under this more stringent denominator strengthens the interpretation that VeriForge's integrated workflow promoted deeper knowledge encoding rather than mere increased exposure. Inter-rater reliability for term coding was good ($\text{ICC}(A,1)=.880$ Baseline, $.790$ VeriForge).
 
\subsubsection{Artifact Evaluation}
 
To address \textbf{RQ4}, two published authors who were not study participants conducted a blind expert rating of all anonymized story snippets using a 7-point Likert scale across three dimensions informed by the verisimilitude literature~\cite{barthes1989reality, wolf2012building, green2000role}: (1) \textit{natural integration of domain details}, whether specialized details feel organically woven into the narrative or read as inserted encyclopedia entries; (2) \textit{specificity over generic tropes}, whether the passage feels grounded and distinctive rather than relying on vague, stock descriptions; and (3) \textit{perceived authorial domain competence}, whether the passage creates a convincing illusion that the author understands the domain. We averaged the two raters' scores for analysis; single-rater absolute-agreement ICCs were moderate ($\text{ICC}(A,1)=0.53\text{--}0.59$). Full study instruments, scale items, and deception experiment details are provided in the supplemental material.

\subsubsection{Statistical Analysis}

We used two-sided paired Wilcoxon signed-rank tests. Zero differences were excluded from each test. Tied absolute differences received average ranks. We report effect sizes as $r=|Z|/\sqrt{N_{\mathrm{eff}}}$. We applied Benjamini--Hochberg FDR correction within three predefined families: item-level questionnaire and behavioral measures, scale composites and subscales, and the three expert-rating dimensions. We used $q<.05$ as the corrected significance threshold. RQ3 and delayed recall analyses were exploratory and were not included in these families. Full details are provided in the supplemental material.

\section{Results}

\begin{figure*}[t!]
  \centering
  \includegraphics[width=\linewidth]{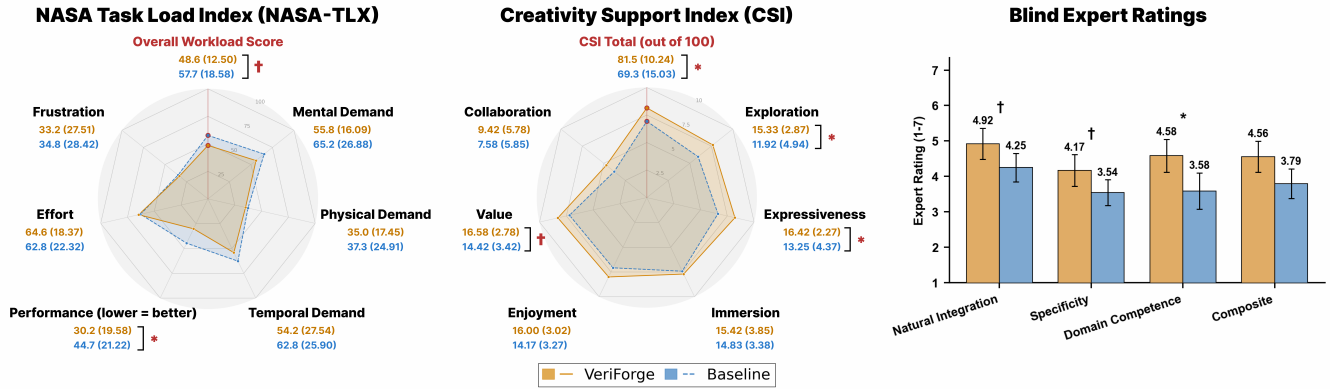}
  \caption{Quantitative results for NASA-TLX (left), CSI (center), and blind expert ratings (right). Values are means (SD). $*$: FDR-corrected $q<.05$; ${\dagger}$: positive but not significant after correction. The expert composite is uncorrected.}
  \Description{A three-panel comparison using orange for VeriForge and blue for Baseline. The NASA-TLX radar shows lower overall workload for VeriForge, 48.6 versus 57.7, with similar effort and lower scores on most other dimensions. The CSI radar shows higher creativity support for VeriForge, 81.5 versus 69.3, with the clearest gains in exploration and expressiveness. Expert-rating bars favor VeriForge on natural integration, specificity, perceived domain competence, and the composite; only domain competence retains a corrected significance marker.}
  \label{fig:results}
\end{figure*}
 
At the aggregate level, VeriForge was associated with higher creativity support (CSI Total: $M=81.47$ vs.\ $M=69.31$; $W=3.0$, $p=.005$, $q=.017$, $r=.804$) and lower overall cognitive workload (NASA-TLX Weighted Total: $M=48.58$ vs.\ $M=57.72$; $W=12.0$, $p=.034$, $q=.060$, $r=.611$), both with large effect sizes (\cref{fig:results}). CSI Total remained significant after FDR correction, while overall workload showed a positive trend that did not survive correction. The overall sensemaking score was also higher under VeriForge ($M=77.25$ vs.\ $M=62.58$; $W=3.0$, $p=.020$, $q=.082$, $r=.770$), but this result did not survive correction. At the subscale level, Insight ($q=.017$) and Connection ($q=.043$) remained significant. SUS usability scores were comparable across conditions ($M=76.46$ vs.\ $M=70.62$; $p=.287$), suggesting that VeriForge's additional features did not compromise perceived ease of use. Our results are organized below around our four primary research questions.
 
\subsection{Illuminating the Unknown: Proactive vs. Reactive Discovery (RQ1)}
 
The most pronounced finding concerned authors' ability to recognize latent knowledge gaps. Participants rated VeriForge's blind-spot alerting significantly higher than the baseline ($M=6.33$ vs.\ $M=3.00$; $W=0.0$, $p=.001$, $q=.016$, $r=.885$), with 11 of 12 scoring VeriForge higher. This result remained significant after FDR correction. In the baseline condition, participants assumed their existing understanding was sufficient and therefore never initiated the searches that would have revealed their gaps. When asked why she had not searched for the distinction between a squire and a full knight, P04 replied: \textit{``Because it never occurred to me.''} P06 described a similar pattern: \textit{``Without the prompts, I wouldn't have noticed the differences between polearm variants... I would have just assumed they were all the same thing.''}
 
Crucially, participants reported that proactive cues did not disrupt writing flow. P06 described the experience as \textit{``essentially frictionless, no window-switching, no context breaks.''} P07 noted that manual research would \textit{``break my writing flow,''} whereas the inline mechanism \textit{``genuinely preserves my writing continuity.''}
 
The confusion reduction item showed the largest absolute shift of any metric in the study ($\Delta=+35.08$; $p=.002$, $q=.016$, $r=.886$). P02 described how a highlight reframed his search strategy: \textit{``It made me realize there are other combat paradigms entirely, so I knew to ask specifically about spear dueling rather than just searching for dueling rules in general.''} The cognitive bridging item showed a positive trend, but did not survive correction ($p=.020$, $q=.082$), while the general gap discovery item did not differ ($p=.313$). One possible interpretation is that proactive highlighting operates more effectively at the level of concrete uncertainty reduction than abstract gap awareness, though the small sample limits conclusions from this null result. This pattern was consistent with a change in exploration behavior: participants issued more queries under VeriForge ($M=6.92$ vs.\ $M=4.00$; $p=.002$, $q=.016$). Canvas nodes and edges also increased (both $q=.016$).
 
\subsection{Scaffolding Sensemaking and Creative Exploration (RQ2)}
 
Creativity support scores were higher under VeriForge. CSI Total ($q=.017$), Exploration ($p=.008$, $q=.022$), and Expressiveness ($p=.004$, $q=.017$) remained significant after correction. Value showed a positive trend, but did not survive correction ($p=.043$, $q=.066$). P01 described the dual-stream panel as complementary: the text stream let knowledge \textit{``sink in more deeply,''} while cards \textit{``let me extract key terms so my brain can work in a more relaxed mode.''}
 
The sensemaking scale provided evidence that the canvas supported relational knowledge building. The overall score was higher under VeriForge ($p=.020$, $q=.082$, $r=.770$), but did not survive correction. Insight ($q=.017$) and Connection ($q=.043$) remained significant, while Structuring was marginal ($q=.051$). Story framework construction showed the strongest raw gain ($\Delta=+19.75$; $p=.008$). P05 described how a system-generated edge between two martial arts schools triggered a plot idea about doctrinal rivalry. P04 compared the canvas to an external scaffold: \textit{``When everything you can use has a structure, it can actually correct your thinking.''}
 
Overall workload was directionally lower under VeriForge, but did not survive correction ($p=.034$, $q=.060$), while Performance remained significant after correction ($p=.002$, $q=.017$). However, Effort remained equivalent ($p=.691$), reflecting cognitive reallocation rather than reduction. P06 articulated this directly: \textit{``The effort was about the same in both, but in one case it went toward creative exploration, and in the other toward the friction of switching windows.''} Immersion did not differ ($p=.578$); participants reported that the canvas competed for attention during high-intensity writing, suggesting a tension between sensemaking depth and writing flow that future work could address.
 
A delayed recall task one week after the study provided exploratory evidence that VeriForge's workflow may promote deeper knowledge internalization. Participants recalled more domain-specific terms from the VeriForge condition ($M=8.04$ vs.\ $M=5.75$; $p=.008$, $r=.891$). Importantly, this difference was not merely a consequence of greater initial exposure: the retention rate, measured as the proportion of domain terms from the original story that reappeared in the recall paragraph, was also higher under VeriForge ($M=.465$ vs.\ $M=.321$; $p=.002$, $r=.883$). Together, these results are consistent with the interpretation that VeriForge's integrated workflow supported not only more extensive domain exploration but also deeper encoding of the knowledge authors incorporated into their writing. These tests were exploratory and were not FDR-corrected.

\subsection{Establishing Trust and Preserving Authorial Agency (RQ3)}
 
RQ3 analyses were exploratory and excluded from FDR correction.

Perceived trust trended higher under VeriForge ($M=5.92$ vs.\ $M=4.75$) but did not reach significance ($p=.125$), likely because both conditions shared the same graph-based retrieval backend with embedded source attributions, elevating trust across conditions and compressing between-condition variance. This suggests that source traceability supports trust as a property of the retrieval infrastructure rather than a differentiator between the two interfaces. However, the deception experiment revealed that this trust operates heuristically rather than through active verification (Section~7.2). Perceived authorial autonomy was identical ($M=6.17$ vs.\ $M=6.17$; $p=1.0$), indicating that VeriForge's proactive assistance did not erode authors' sense of creative control.
 
The controlled deception experiment offered a more nuanced picture. Only 1 of 12 participants detected the source-card mismatch. The remaining 11 did not notice. This test measured whether participants inspected the linked evidence. It did not test whether they could identify a false domain claim from their own knowledge. P08 noted that when a topic sparked his interest, the source label on the card was easy to locate without rereading the full response, yet this accessibility facilitated rapid heuristic consumption rather than careful verification. P06 acknowledged being deceived but rationalized that a system producing well-formatted, source-linked cards was unlikely to be wrong, so he \textit{``chose to believe it.''} P07 was more direct: \textit{``I would believe everything it gives me.''} We note that under normal operating conditions this trust is well-placed, as all Knowledge Cards are derived from user-uploaded primary sources; the broader implications of this provenance-as-authority effect are discussed in Section~7.2.
 
Despite this trust in factual outputs, participants drew a sharp boundary around narrative synthesis. P07 rejected AI-generated prose from past experience but valued VeriForge's concrete Knowledge Cards: \textit{``Give me a specific sequence of moves... that's what becomes a highlight in my writing.''} P05 used the system exclusively for technique names and historical facts but refused to let it touch his combat scenes. These responses are consistent with the view that presenting domain knowledge as structured raw material, rather than finished prose, may help preserve authorial ownership, though this interpretation draws on participants' retrospective comparisons with prior tools rather than a controlled manipulation within our study.

\subsection{Domain Grounding in Resulting Passages (RQ4)}
 
Blind expert ratings provided preliminary evidence that VeriForge narratives received higher scores across all three evaluation dimensions (\cref{fig:results}). Only perceived authorial domain competence remained significant after correction ($M=4.58$ vs.\ $M=3.58$; $W=2.5$, $p=.005$, $q=.015$, $r=.818$). Natural integration of domain details showed a positive trend that did not survive correction ($M=4.92$ vs.\ $M=4.25$; $p=.045$, $q=.065$, $r=.617$), as did specificity ($M=4.17$ vs.\ $M=3.54$; $p=.065$, $q=.065$, $r=.536$). The composite score was higher for VeriForge ($M=4.56$ vs.\ $M=3.79$; $p=.005$, $r=.770$), with 10 of 12 pairs favoring the system; this composite test was uncorrected. While inter-rater reliability was moderate ($\text{ICC}(A,1) = 0.53\text{--}0.59$), both raters independently assigned higher scores to VeriForge snippets in these same 10 pairs, suggesting a consistent directional signal despite scoring variance. Moderate ICC is expected in literary evaluation given the inherent subjectivity of assessing stylistic nuance~\cite{Kaufman,Lumley}.
 
Word count did not differ significantly between conditions ($M=285.83$ vs.\ $M=258.25$; $p=.532$), providing no evidence that the additional sensemaking activity in VeriForge reduced writing productivity. Taken together, these findings offer early-stage evidence that VeriForge's mixed-initiative workflow may be associated with improvements in perceived prose quality, particularly in conveying a convincing sense of domain familiarity, without slowing authors down. Whether these effects hold over longer, more developed passages remains an important question for future work.

\section{Discussion}

\subsection{Beyond Creative Writing: Generalizing the Discovery-Synthesis Split}

While VeriForge targets fiction writers, the underlying design principle extends to any knowledge work where the final product must carry a unique human perspective. The discovery-synthesis split is most valuable precisely where generative AI is most dangerous: in work where voice, interpretation, and judgment constitute the output's core value. In journalism, reporters routinely rely on domain experts as background knowledge providers while retaining full ownership of the narrative frame~\cite{albaek2011interaction}, a human enactment of the same split VeriForge automates. In documentary filmmaking, cinematic artistry without genuine domain understanding produces representations that obscure rather than illuminate their subjects~\cite{ruby2000picturing}. Writing-to-learn research has long demonstrated that transforming unfamiliar knowledge into one's own prose is itself the mechanism through which deeper understanding forms~\cite{delapaz2005effects, newell2006writing}, reinforcing the knowledge-transforming model~\cite{bereiter1987psychology} that underpins our design. Across these domains, the same pattern holds: practitioners who lack domain familiarity face blind spots they cannot self-diagnose, and the cognitive work of synthesis is not a bottleneck to be optimized away but the locus of the product's value.
 
That said, generalizing the split is not plug-and-play. A journalist working against a deadline has a far narrower window for knowledge discovery than a novelist drafting over months; an academic writing within their own specialty faces a much shallower IOED than a fiction writer entering an unfamiliar domain. These variations suggest that the discovery-synthesis split is best understood as a design \textit{framework} whose parameters, including the aggressiveness of proactive cues, the granularity of knowledge scaffolding, and the degree of source traceability, must be calibrated to the epistemic conditions of each domain.                                                             

\subsection{The Provenance Paradox}
 
Our deception experiment carries implications beyond creative writing. Only 1 of 12 participants detected the source-card mismatch. The remaining 11 accepted it without hesitation. P06 rationalized that a system producing well-formatted, source-linked cards was unlikely to be wrong. The source label functioned not as an invitation to verify but as a seal of authority that reduced the perceived need for verification. This result concerns evidence inspection, not the ability to identify false domain claims. An incorrect AI-generated description would test hallucination detection and may depend on prior expertise. Future work should pair an incorrect description with a correct source passage to study that question separately.
 
This behavior is well explained by dual-process accounts of persuasion~\cite{petty1986elaboration}: when users lack both the ability and the motivation to scrutinize a claim, they fall back on surface cues, such as whether a citation looks authoritative, rather than engaging with its substance. Interface design amplifies this tendency, as structured formatting and institutional labels serve as trust heuristics in their own right~\cite{sundar2008main}. Combined with IOED, this creates a double bind: authors cannot recognize \textit{what} to verify, and peripheral cues remove the impulse to verify \textit{at all}. Provenance, intended to support transparency, instead short-circuits the epistemic vigilance it was designed to enable.
 
This poses a design tension that sits at the heart of VeriForge's paradigm. The system is engineered to protect creative flow; yet that very frictionlessness is what enables uncritical acceptance. Adding heavier verification mechanisms risks destroying the flow the system was built to preserve. Research on intelligibility demand~\cite{lim2009assessing} suggests a way forward: users' need for transparency is context-dependent, rising with perceived consequences. Provenance design should therefore be \textit{risk-adaptive}, calibrating epistemic friction to the stakes of the task. In creative writing, lightweight cues may suffice; in legal or clinical retrieval, users may need interactive mechanisms that invite active inspection rather than passive consumption~\cite{kulesza2015principles}. How to design such mechanisms without disrupting workflow remains an open and increasingly urgent challenge as retrieval-augmented systems move into higher-stakes domains.
 
\subsection{Limitations and Future Work}

\textbf{Task and evaluation scope.} We used 30-minute sessions to isolate the cold-start phase of domain discovery. This duration aligns with prior AI writing studies~\cite{fu2026vistoria,talaei2025storysage,suh2024luminate}, but captures only an immediate knowledge-transforming loop. We therefore treat the RQ4 expert ratings as initial evidence of domain grounding. Longitudinal studies should test longer passages and sustained use. Participants also noted that the Knowledge Canvas can become visually dense. Future work should explore semantic clustering and context-based filtering.
 
\textbf{Genre and sample scope.} Our summative study fixed the genre to historical martial scenes, a dense, source-verifiable setting suited to observing gap discovery and factual grounding in a short session. This bounds where VeriForge has been shown to work, not where the problem exists. Formative writers had latent gaps across folklore, science fiction, crime fiction, historical romance, and cosmic horror, including gaps unrelated to historical periods. Post-study comments also suggested transfer: two participants mentioned science fiction, and one emphasized detail-rich subjects. Cross-genre evaluation remains future work, especially in domains with different knowledge structures, such as legal, medical, and other professional settings. Our sample ($N=12$, age $M=21.92$) consisted of hobbyists and online novelists with moderate dedicated writing experience ($M=3.18$ years), representing writers most likely to encounter IOED-driven blind spots. Professional novelists with deeply internalized domain expertise may have fewer latent gaps, potentially reducing the benefit of proactive highlighting.

Additional evaluation limitations are provided in Appendix~\ref{app:evaluation-reliability}.

\section{Conclusion}

This paper presents VeriForge, a mixed-initiative writing environment that supports fiction writers in discovering and integrating unfamiliar domain knowledge. Grounded in formative interviews with nine writers, VeriForge separates knowledge discovery from narrative synthesis. Proactive highlights surface potential gaps during drafting. Source-anchored Knowledge Cards pair an AI-generated description with a quoted source passage, while a spatial Knowledge Canvas helps writers organize and connect selected knowledge without giving up control of the prose. Through a within-subjects study with 12 writers, we found that VeriForge helped participants notice overlooked gaps, explore domain details, and produce passages with stronger perceived domain grounding. Participants also preserved their sense of agency by treating retrieved knowledge as material for writing rather than finished text. At the same time, our deception study showed that visible provenance can become an authority cue when users do not inspect the source. We hope VeriForge points toward AI writing tools that help authors see what they do not yet know while keeping interpretation and narrative decisions in human hands.

\begin{acks}
We acknowledge the partial use of a large language model (LLM), specifically ChatGPT, to assist in the writing process. The LLM was employed as a tool for polishing the manuscript to enhance the clarity and quality of the text.
\end{acks}

\bibliographystyle{ACM-Reference-Format}
\bibliography{sample-base}

\appendix
\onecolumn
\section{Participant Demographics}
\label{app:demographics}

\begin{table}[h]
\centering
\caption{Demographics of formative study participants ($N=9$).}
\label{tab:formative-demographics}
\footnotesize
\renewcommand{\arraystretch}{1.15}
\setlength{\tabcolsep}{4pt}
\begin{tabular*}{\textwidth}{@{\extracolsep{\fill}}l c c l l p{0.38\textwidth} l@{}}
\toprule
ID & Age & Gender & Education & Status & Primary genre and practice & AI tools \\
\midrule
F01 & 22 & F & Undergrad & Hobbyist & Cosmic horror; posts to online communities & DeepSeek \\
F02 & 24 & F & Master's & Hobbyist & Romance / Boys' Love; posts to online communities & Doubao \\
F03 & 26 & F & PhD & Professional & Women's romance; published author, contracted, radio-drama adaptation & Gemini \\
F04 & 20 & M & Undergrad & Professional & Science fiction; published in SF magazines & ChatGPT, Gemini \\
F05 & 19 & F & Undergrad & Hobbyist & Women-oriented detective/crime fiction; member of literary society & ChatGPT \\
F06 & 19 & M & Undergrad & Hobbyist & Science fiction; SF and literary societies, writing competitions & ChatGPT, Claude \\
F07 & 30 & F & Undergrad & Professional & Historical romance; contracted, film/TV adaptation deal & ChatGPT, Gemini \\
F08 & 27 & F & Associate & Professional & Period fiction (20th-c. China); contracted, 1M+ platform views & DeepSeek \\
F09 & 28 & F & Undergrad & Professional & Folklore / ethnic horror; contracted, radio adaptation, 100k+ readers & DeepSeek, Gemini, ChatGPT \\
\bottomrule
\end{tabular*}
\end{table}

\begin{table}[h]
\centering
\caption{Demographics of user study participants ($N=12$).}
\label{tab:user-demographics}
\footnotesize
\renewcommand{\arraystretch}{1.15}
\setlength{\tabcolsep}{8pt}
\begin{tabular*}{\textwidth}{@{\extracolsep{\fill}}l c c l r l@{}}
\toprule
ID & Age & Gender & Education & Writing exp. (yrs) & Writer type \\
\midrule
P01 & 20 & M & Undergrad & 1 & Hobbyist \\
P02 & 20 & M & Undergrad & 3 & Hobbyist \\
P03 & 26 & F & Undergrad & 8 & Online novelist \\
P04 & 22 & M & Master's & 1 & Hobbyist \\
P05 & 20 & M & Undergrad & 1 & Hobbyist \\
P06 & 21 & F & Undergrad & 1 & Hobbyist \\
P07 & 22 & M & Master's & 2 & Hobbyist \\
P08 & 24 & F & Undergrad & 4 & Online novelist \\
P09 & 24 & F & Master's & 4 & Hobbyist \\
P10 & 21 & F & Undergrad & 1 & Hobbyist \\
P11 & 22 & F & Master's & 7 & Online novelist \\
P12 & 21 & F & Undergrad & 2 & Online novelist \\
\bottomrule
\end{tabular*}
\end{table}

\section{Baseline Feature Comparison}
\label{app:baseline-comparison}

\begin{table}[h]
\centering
\caption{Feature comparison of Baseline and VeriForge under shared retrieval infrastructure.}
\label{tab:baseline-comparison}
\footnotesize
\setlength{\tabcolsep}{8pt}
\begin{tabular*}{\textwidth}{@{\extracolsep{\fill}}p{0.78\textwidth}cc@{}}
\toprule
Feature & Baseline & VeriForge \\
\midrule
Text editor for drafting & Yes & Yes \\
Conversational query panel & Yes & Yes \\
Miro-like spatial canvas & Yes & Yes \\
Graph-based retrieval backend & Yes & Yes \\
Shared Semantic Frame & Yes & Yes \\
Source-grounded responses & Yes & Yes \\
Proactive inline highlighting & No & Yes \\
Highlight-derived Knowledge Cards & No & Yes \\
Dual-stream text-card coupling & No & Yes \\
Gold canvas nodes from editor highlights & No & Yes \\
AI-reasoned canvas connections & No & Yes \\
\bottomrule
\end{tabular*}
\end{table}

\section{Evaluation Reliability}
\label{app:evaluation-reliability}

Our strengthened baseline shares the same graph-based retrieval backend as VeriForge, ensuring that observed differences reflect interaction design rather than retrieval quality. However, a future comparison against dedicated knowledge management tools such as Obsidian or NotebookLM would further situate VeriForge in the broader tool landscape, though we note that such tools require considerable technical overhead and are rarely adopted by creative writers in practice. Inter-rater agreement for the blind expert evaluation was moderate ($\text{ICC}(A,1) = 0.53\text{--}0.59$), reflecting the inherent subjectivity of literary quality assessment; future studies could employ larger rater pools or more granular rubric anchoring to improve reliability.

\end{document}